\documentstyle{europhys}
\def\And{{\rm and\ }}

\def\stars{\bigskip\centerline{***}\medskip}

\newif\ifboo \boofalse

\def\Review#1{\boofalse{\it #1},}
\def\Name#1{{\sc #1},}
\def\Vol#1{\ifboo Vol. {\bf #1}\else{\bf #1}\fi}
\def\Year#1{\ifboo #1\else(#1)\fi}

\def\Page#1{\ifboo {\rm p. #1}\else{\rm #1}\fi}
\begin{document}
%
%%%   The headers.
%
%%%   These three macros are to have correct headings in your paper.
%%%   You shall omit all the arguments in the two macros `\euro{}{}{}{}'
%%%   `\Date{}' and fill in `\shorttitle{}'. 
%%%   If there is more than one author in the 
%%%   \shorttitle macro, use the macro \etal after first author's name
%%%   to obtain the correct heading.
%
\euro{?}{?}{?}{2000}
\Date{version of 15/ 2 / 2000}
\shorttitle{W. WERNSDORFER {\it et al.} LANDAU ZENER TUNNELING ETC.}
\title{Nonadiabatic Landau Zener tunneling in Fe$_8$ molecular nanomagnets}
\author{W. Wernsdorfer\inst{1}, R. Sessoli\inst{2}, A. Caneschi\inst{2}, 
D. Gatteschi\inst{2} \And A. Cornia\inst{3}}
\institute{
     \inst{1} Lab. de Mag. Louis N\'eel $-$ CNRS, BP166, 38042 Grenoble, France\\
     \inst{2} Dept. of Chemistry, Univ. of Florence, 
                       Via Maragliano 77, 50144 Firenze, Italy\\
     \inst{3} Dept. of Chemistry, Univ. of Modena, 
                       Via G. Campi 183, 41100 Modena, Italy}
\rec{28 Oct. 99}{}
\pacs{
\Pacs{75}{45$+$j}{Macroscopic quantum phenomena in magnetic systems}
\Pacs{75}{50 Tt}{Fine-particle systems}
\Pacs{75}{60 Ej }{Magnetisation curves, hysteresis, Barkhausen and related effects}
      }
\maketitle
\begin{abstract}
The Landau Zener method allows to measure very small tunnel splittings $\Delta$ 
in molecular clusters Fe$_8$. The observed oscillations 
of $\Delta$ as a function of the magnetic field applied along the hard anisotropy axis are 
explained in terms of topological quantum interference of two tunnel paths of opposite 
windings. Studies of the temperature dependence of the Landau Zener transition rate $P$
gives access to the topological quantum interference between excited spin levels.
The influence of nuclear spins is demonstrated by comparing $P$ of the 
standard Fe$_8$ sample with two isotopically substituted samples. 
The need of a generalised Landau Zener transition rate theory is shown.
\end{abstract}
%
%%%   Main text
%%%   Sectioning
%%%   In EuroPhys there is only ``one'' level of sectioning `\section{}'.
%
During the last few decades, a large effort has been spent to understand the detailed dynamics of quantum systems that are exposed to time-dependent external fields and dissipative effects \cite{Grifoni98}. It has been shown that molecular magnets offer an unique opportunity to explore the quantum dynamics of a large but finite spin. 
These molecules are the final point in the series of smaller and smaller units 
from bulk magnets to single magnetic moments. They are regularly 
assembled in large crystals where often all molecules have the same orientation. Hence, 
macroscopic measurements can give direct access to single molecule properties.

The most 
prominent examples are a dodecanuclear mixed-valence manganese-oxo cluster with 
acetate ligands, Mn$_{12}$~\cite{Sessoli93}, and an octanuclear iron(III) oxo-
hydroxo cluster of formula [Fe$_8$O$_2$(OH)$_{12}$(tacn)$_6$]$^{8+}$, 
Fe$_8$~\cite{Barra96}, 
where tacn is a macrocyclic ligand. Both systems have a spin 
ground state of $S = 10$, and an Ising-type magneto-crystalline anisotropy, which 
stabilises the spin states with the quantum numbers 
$M = \pm10$ and generates an energy barrier for the 
reversal of the magnetisation of about 67~K for Mn$_{12}$ and 25~K for 
Fe$_8$~\cite{Sessoli93,Barra96}.

Fe$_8$ is particular interesting for studies of quantum tunnelling because it shows a pure quantum regime, i.e. below 360 mK the relaxation is purely due to quantum
tunnelling, and not to thermal activation \cite{Sangregorio97}. We showed recently that the Landau Zener method can be used to measure the very small tunnel splittings $\Delta$ 
in Fe$_8$ \cite{Science99}. The observed oscillations 
of $\Delta$ as a function of the magnetic field applied along the hard anisotropy axis are 
explained in terms of topological quantum interference of two tunnel paths of opposite 
windings which was predicted by Garg \cite{Garg93}. 
This observation was the first direct evidence of the topological part of the 
quantum spin phase (Berry or Haldane phase \cite{Berry,Haldane}) in a magnetic system.

Recently, we demonstrate the influence of nuclear spins, proposed by
Prokof'ev and Stamp~\cite{Prok_Stamp98}, by comparing
relaxation and hole digging measurements \cite{PRL_Fe57}
of two isotopically substituted samples: 
(i) the hyperfine coupling was increased by the 
substitution of $^{56}$Fe with $^{57}$Fe, 
and (ii) decreased by the substitution of $^1$H with $^2$H.
These measurements were supported quantitatively
by numerical simulations taking into account the altered 
hyperfine coupling \cite{PRL_Fe57,Rose00}. 

In this letter, we present studies of the temperature dependence of the 
Landau Zener transition rate $P$
yielding a deeper insight into the spin dynamics of the Fe$_8$ cluster. 
By comparing the three isotopic samples we confirm the influence
of nuclear spins on the tunneling mechanism and in particular on the
lifetime of the first excited states.
Our measurements show the need of 
a generalised Landau Zener transition rate theory taking into account environmental 
effects such as hyperfine and spin--phonon coupling \cite{Igor_review}.

All measurements of this article were performed using a new technique of 
micro-SQUIDs where the sample is directly coupled with an array of 
micro-SQUIDs \cite{WW_PhD}. The high sensitivity of this magnetometer allows us to study single Fe$_8$ crystals 
\cite{Wieghardt} of the order of 10 to 500 $\mu$m. 

The crystals of the standard Fe8 cluster, $^{\rm st}$Fe$_8$ or Fe$_8$,
[Fe$_8$(tacn)$_6$O$_2$(OH)$_{12}$]Br$_8$.9H$_2$O
where tacn = 1,4,7- triazacyclononane, were prepared as reported by
Wieghardt {\it et al.}~\cite{Wieghardt}. 
For the synthesis of the
$^{57}$Fe-enriched sample, $^{57}$Fe$_8$, 
a 13 mg foil of 95$\%$ enriched $^{57}$Fe 
was dissolved in a few drops of
HCl/HNO$_3$ (3 : 1) and the resulting solution was used as the iron
source in the standard procedure. The $^2$H-enriched Fe$_8$ sample,
$^{\rm D}$Fe$_8$, was crystallised from pyridine-d$_5$ and D$_2$O (99$\%$)
under an inert atmosphere at 5$^{\circ}$C by using a non-deuterated
Fe(tacn)Cl$_3$ precursor. The amount of isotope exchange was not
quantitatively evaluated, but it can be reasonably assumed that the H
atoms of H$_2$O and of the bridging OH groups, as well as a part
of those of the NH groups of the tacn ligands are replaced by deuterium
while the aliphatic hydrogens are essentially not affected. The crystalline
materials were carefully checked by elemental analysis and
single-crystal X-ray diffraction.

\begin{figure}
\centerline{\epsfxsize=8 cm \epsfbox{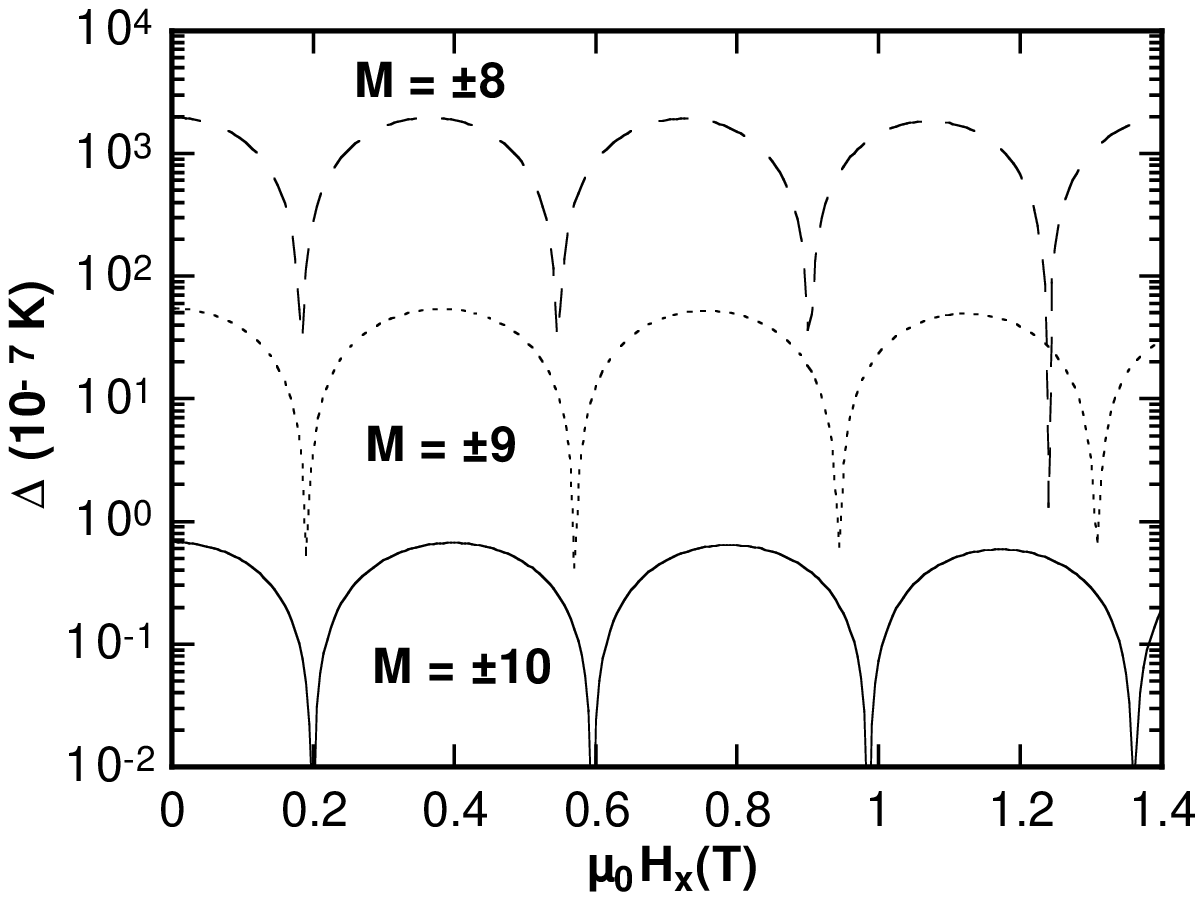}}
\caption{Calculated tunnel splitting $\Delta_{M,M'}$ (Eq.~\ref{eq_H_biax}) 
as a function of the transverse field $H_x$ for 
quantum transition between $M = \pm10, \pm9$ and $\pm8$.}
\label{Delta_10_9_8}
\end{figure}

The simplest model describing the spin system of Fe$_8$ molecular clusters has the 
following Hamiltonian \cite{Barra96}:
\begin{equation}
H = D S_z^2 + E \left(S_x^2 - S_y^2\right) + H_2 - g \mu_B \mu_0 \vec{S}\vec{H}
\label{eq_H_biax}
\end{equation}
$S_x$, $S_y$, and $S_z$ are the three components of the spin operator, $D$  and $E$ 
are the anisotropy constants, $H_2$ takes into account weak higher order terms \cite{Caciuffo,Barra_new}, and the last term of the Hamiltonian describes the Zeeman 
energy associated with an applied field $\vec{H}$. 
This Hamiltonian defines a hard, medium, 
and easy axes of magnetisation in $x$, $y$ and $z$ direction, 
respectively. It 
has an energy level spectrum with $(2S+1) = 21$ values which, in first approximation, 
can be labelled by the quantum numbers $M~= -10, -9, ...10$. The energy spectrum, can be obtained by using 
standard diagonalisation techniques of the $[21 \times 21]$ matrix describing the spin 
Hamiltonian $S = 10$. 
At $\vec{H} = 0$, the levels $M = \pm10$ have the lowest energy. 
When a field $H_z$ is applied, the energy levels with $M < < 0$ increase, 
while those with $M >> 0$ decrease. Therefore, different energy values can cross at certain fields. This crossing can be avoided by transverse terms containing $S_x$ or 
$S_y$ spin operators which split the levels. The spin $S$ is {\it in resonance} 
between two states $M$ and $M'$ when the local longitudinal field 
is close to such an avoided energy level crossing ($|H_z| < 10^{-8}$ T 
for the avoided level crossing around $H_z$ = 0). 
The energy gap, the so-called 
tunnel spitting $\Delta_{M,M'}$, can be tuned by an applied field in the $xy-$plane 
via the $S_xH_x$ and $S_yH_y$ Zeeman terms. It turns out that a 
field in $H_x$ direction (hard anisotropy direction) can periodically change the tunnel 
spitting $\Delta$ as displayed in Fig.~\ref{Delta_10_9_8} where $H_2$ in Eq.~\ref{eq_H_biax} was taken from \cite{Barra_new}. 
In a semi-classical description, these oscillations are due to constructive 
or destructive interference of quantum spin phases of two tunnel paths \cite{Garg93}.

A direct way of measuring the tunnel splittings $\Delta_{M,M'}$ is by using the Landau-
Zener model \cite{Zener,Zener_new} which gives the tunnelling probability 
$P_{M,M'}$ when sweeping the 
longitudinal field $H_z$ at a constant rate over an avoided energy level 
crossing \cite{remark_Loss}:
\begin{equation}
P_{M,M'} = 1 - e^{-
\frac {\pi \Delta_{M,M'}^2}{2 \hbar g \mu_B |M - M'| \mu_0 dH_z/dt}}
\label{eq_LZ}
\end{equation}
Here, $M$ and $M'$ are the quantum numbers of the 
avoided energy level crossing, $dH_z/dt$ is the 
constant field sweeping rate, $g~\approx~2$, 
$\mu_B$ the Bohr magneton, and $\hbar$ is Planck's constant. 

\begin{figure}
\centerline{\epsfxsize=8 cm \epsfbox{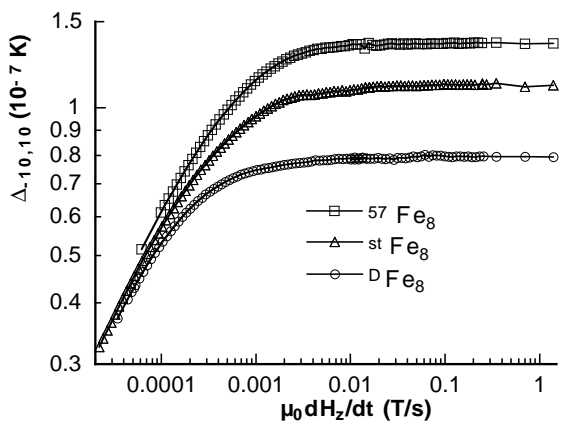}}
\caption{Field sweeping rate dependence of the tunnel splitting $\Delta_{-10,10}$ 
measured by a Landau Zener method for three Fe$_8$ samples, for $H_x = 0$. 
The Landau Zener method
works in the region of high sweeping rates where 
$\Delta_{-10,10}$ is sweeping rate independent. 
Note that the differences of $\Delta_{-10,10}$ between the three samples are rather small in comparison to the oscillations in Fig.~\ref{P_Hx_T}.}
\label{LZ_test}
\end{figure}

In order to apply the Landau-Zener formula (Eq.~\ref{eq_LZ}), we first cooled the 
sample from 5 K down to 0.04 K in a field of $H_z$~=~-1.4~T yielding a negative 
saturated magnetisation state.
Then, we swept the applied field at a constant rate 
over the zero field resonance transition and measured the fraction of molecules which 
reversed their spin. This procedure yields the tunnelling 
rate $P_{-10,10}$ and thus the tunnel 
splitting $\Delta_{-10,10}$ (Eq.~\ref{eq_LZ}). The predicted Landau-Zener 
sweeping field dependence of $P_{-10,10}$ can be checked by 
plotting $\Delta_{-10,10}$ as a function of the field sweeping rate 
which should show a constant which was indeed the 
case for sweeping rates between 1 and 0.001~T/s (fig.~\ref{LZ_test}). The deviations 
at lower sweeping rates are mainly due to the 
{\it hole-digging mechanism}~\cite{Ohm,PRL_dig} 
which slows down the relaxation.
The comparison with the isotopically substituted Fe$_8$ samples shows a 
clear dependence of $\Delta_{-10,10}$ on the hyperfine coupling (Fig.~\ref{LZ_test}). Such an effect has been predicted for a {\it constant} applied field by Tupitsyn {\it et al.} \cite{Tupitsyn97}.

All measurement so far were done in the pure quantum regime 
($T < 0.36$ K) where transition via excited spin levels can be neglected. 
We discuss now the temperature region of small thermal activation ($T < 0.7$ K) 
where we should consider transition via excited spin levels \cite{Fort98}. 
We make the Ansatz 
that only ground state tunnelling ($M = \pm10$) and transitions via the first 
excited spin levels ($M = \pm9$) are relevant for temperatures slightly 
above 0.36 K. We will see that this Ansatz describes well our 
experimental data but, nevertheless, it 
would be important to work out a complete theory \cite{Leuenberger00}. 

In order to measure the temperature dependence of the transition rate, 
we used the Landau--Zener method~\cite{Zener} 
as described above with a phenomenological 
modification of the transition rate $P$ (for a negative saturated 
magnetisation):
\begin{equation}
P = n_{-10}P_{-10,10} + P_{th}
\label{eq_LZ_T}
\end{equation}
where $P_{-10,10}$ is given by Eq.~\ref{eq_LZ} , $n_{-10}$ is the 
Boltzmann population of the $M = -10$ spin level, and 
$P_{th}$ is the overall transition rate via excited spin levels.
$n_{-10} \approx 1$ for the considered temperature $T < 0.7$ K 
and a negative saturated magnetisation of the sample.

\begin{figure}
\centerline{\epsfxsize=8 cm \epsfbox{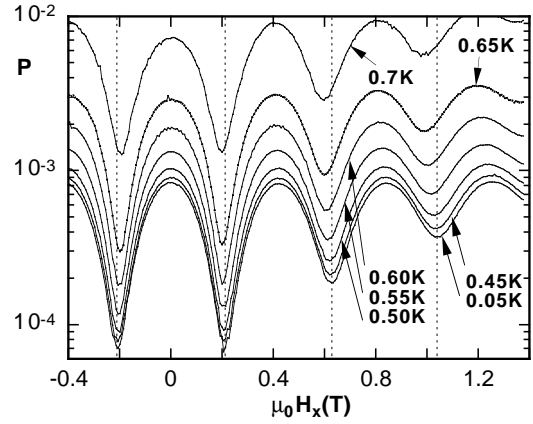}}
\caption{Transverse field dependence of the transition rate $P$ at several 
temperatures, and the ground state transition rate $P_{-10,10}$ measured 
at $T = 0.05$ K and for $^{st}$Fe$_8$. The field sweeping rate was 0.14 T/s. 
The dotted lines indicate the minima of $P_{-10,10}$.}
\label{P_Hx_T}
\end{figure}

\begin{figure}
\centerline{\epsfxsize=8 cm \epsfbox{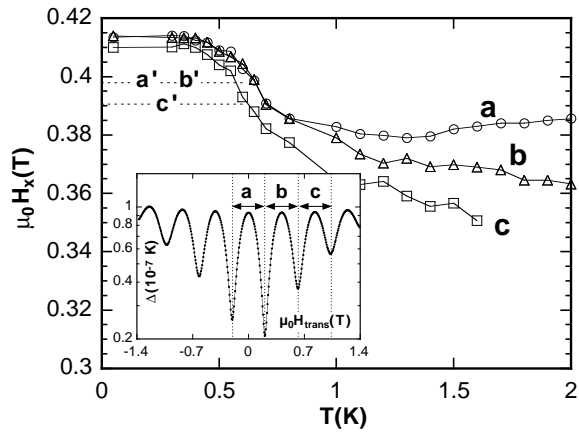}}
\caption{Temperature dependence of the field between minima of 
the transition rate $P$ in Fig.~\ref{P_Hx_T}. 
a, b, and c are defined in the inset. 
The dotted line labelled with a', b', and c' where take from 
$P_{th}$ of Fig.~\ref{P_th_Hx_T}; see also \cite{Berry_Hall}.}
\label{period_T}
\end{figure}

Fig.~\ref{P_Hx_T} displays the measured transition rate $P$ 
for $^{st}$Fe$_8$ as a 
function of a transverse field $H_x$ and for several temperatures. 
The oscillation of $P$ are seen for all temperatures but the periods 
of oscillations decreases for increasing temperature 
(Fig.~\ref{period_T}). This behaviour can be explained by the giant 
spin model (Eq.~\ref{eq_H_biax}) with higher order 
transverse terms ($H_2$). Indeed, the tunnel 
splittings of excited spin levels oscillate as a function of $H_x$ with 
decreasing periods (Fig.~\ref{Delta_10_9_8}).

\begin{figure}
\centerline{\epsfxsize=8 cm \epsfbox{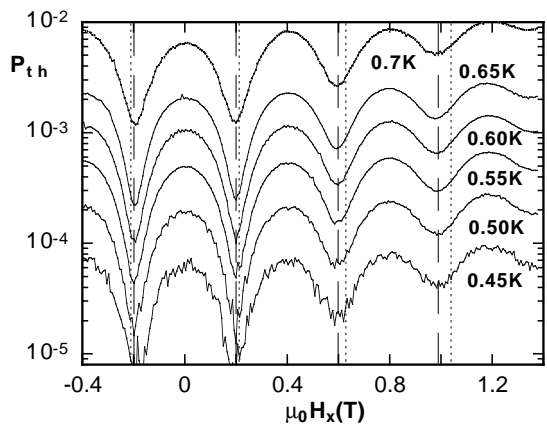}}
\caption{Transverse field dependence of $P_{th}$ which is the difference 
between the measures tunnel probability $P$ and the ground state tunnel probability $n_{-10}P_{-10,10}$ measured at T = 0.05K (see Fig.~\ref{P_Hx_T}. 
The field sweeping rate was 0.14 T/s. The long dotted lines indicate the minima of $P_{th}$ whereas the short dotted lines indicate the minima of $P_{-10,10}$.}
\label{P_th_Hx_T}
\end{figure}

Fig.~\ref{P_th_Hx_T} displays the transition rate via excited spin levels
$P_{th} = P - n_{-10}P_{-10,10}$. Surprisingly, the periods of 
$P_{th}$ are temperature independent in the region $T < 0.7$ K. 
This suggests that only transitions via excited levels $M = \pm9$ are 
important in this temperature regime. This statement is confirmed by 
the following estimation \cite{remark_villain}, 
see also Ref. \cite{Leuenberger00}.

Using Eq.~\ref{eq_LZ}, typical field sweeping rates of 0.1 T/s, and 
tunnel splittings from Fig.~\ref{Delta_10_9_8}, one easily finds that the Landau Zener 
transition probability of excited levels are $P_{-M,M} \approx 1$ for 
$M < 10$ and $\vec{H} \approx 0$. This means that the relaxation 
rates via excited levels are 
mainly governed by the lifetime of the excited levels and the time 
$\tau_{res,M}$ during which these levels are in resonance. 
The later can be estimated by 
\begin{equation}
\tau_{res,M} = \frac{\Delta_{-M,M}}{g \mu_B M \mu_0 dH_z/dt}.
\label{eq_tau_res}
\end{equation}

The probability for a spin to pass into the excited level $M$ 
can be estimated by $\tau_M^{-1}e^{-E_{10,M}/k_BT}$, where $E_{10,M}$ 
is the energy gap between the levels $10$ and $M$, and $\tau_M$ is 
the lifetime of the excited level $M$. We yield \cite{remark_villain}
\begin{equation}
P_{th} \approx \sum_{M = 9, 8} \frac{\tau_{res,M}}{\tau_M}e^{-E_{10,M}/k_BT} 
\approx \sum_{M = 9, 8} \frac{\Delta_{-M,M}}{\tau_M g \mu_B 
M \mu_0 dH_z/dt}e^{-E_{10,M}/k_BT}.
\label{eq_P_th_9}
\end{equation}

\begin{figure}
\centerline{\epsfxsize=8 cm \epsfbox{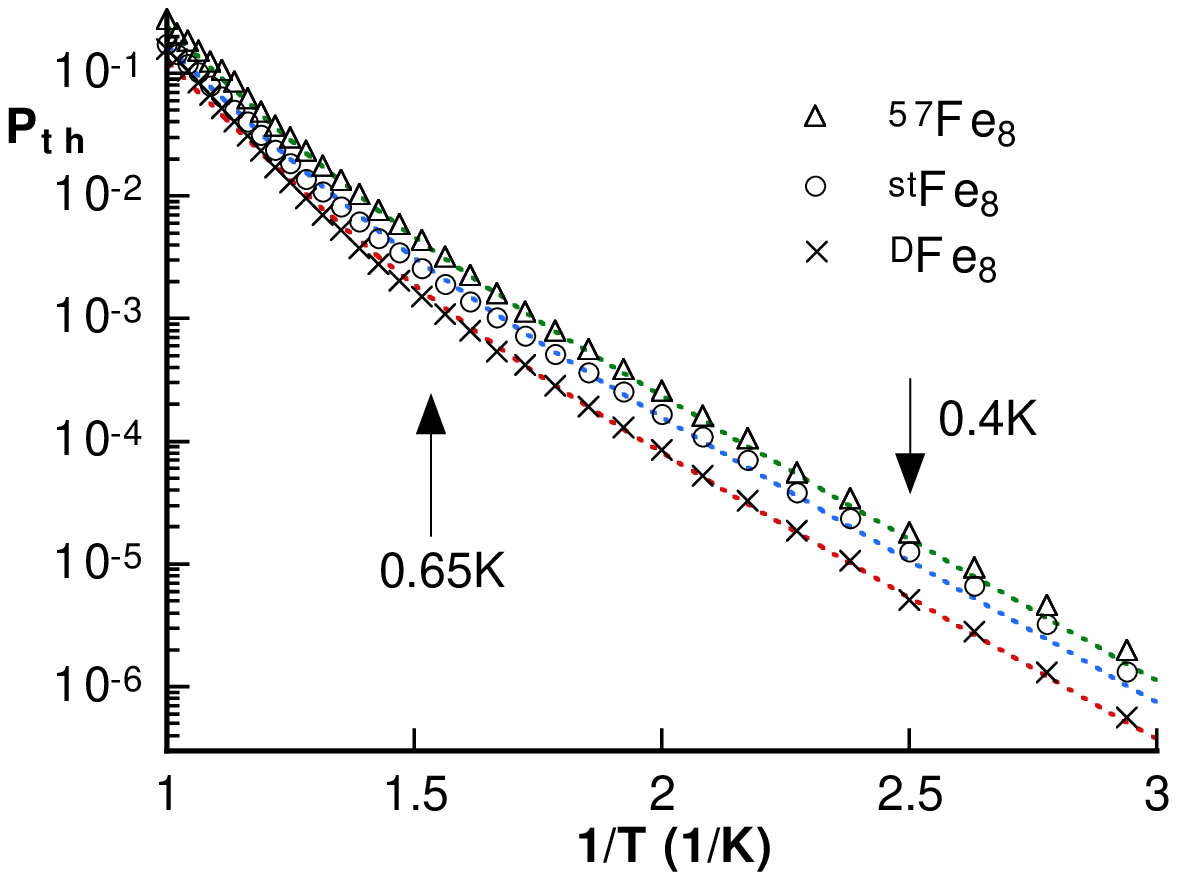}}
\caption{Temperature dependences of $P_{th}$ for $H_x$ = 0 
for three Fe$_8$ samples. The field sweeping rate was 0.14 T/s. 
The dotted lines are fits of the data using Eq.~\ref{eq_P_th_9} \cite{remark_const}.}
\label{P_th_T}
\end{figure}

Note that this estimation neglects higher excited levels with $|M| < 8$ 
\cite{remark_levels}.
Fig.~\ref{P_th_T} displays the measured $P_{th}$ for the 
three isotopic Fe$_8$ samples. 
For 0.4 K $< T <$ 1 K we fitted Eq.~\ref{eq_P_th_9} to the data leaving only
the level lifetimes $\tau_9$ and $\tau_8$ as adjustable parameters. All other
parameters are calculated using Eq.~\ref{eq_H_biax} \cite{remark_const}. 
We obtain $\tau_9 = 1.0, 0.5,$ and $0.3 \times10^{-6}$s, and 
$\tau_8 = 0.7, 0.5,$ and $0.4 \times10^{-7}$s 
for $^{D}$Fe$_8$, $^{st}$Fe$_8$, and $^{57}$Fe$_8$, respectively. 
This result justifies our Ansatz of considering only the first excited 
level for 0.4 K $< T <$ 0.7 K. Indeed, the second term of the summation in 
Eq.~\ref{eq_P_th_9} is negligible in this temperature interval.
It is interesting to note that this finding is in contrast to hysteresis loop 
measurements on Mn$_{12}$ \cite{Kent00} which suggested an abrupt 
transition between thermal assisted and pure quantum tunnelling \cite{Garanin}.
Furthermore, our result shows clearly the influence of
nuclear spins which seem to decrease the level lifetimes $\tau_M$,
{\it i.e.} to increase dissipative effects.

The nuclear magnetic moment and not the mass of the nuclei 
seems to have the major effect on the dynamics of the
magnetization. In fact the mass is increased in both isotopically modified
samples whereas the effect on the the relaxation rate is opposite. 
On the other hand ac susceptibility measurements at
$T~>$~1.5~K showed no clear difference between
the three samples \cite{Paulsen} suggesting that
above this temperature, where the relaxation is predominately due to
spin-phonon coupling~\cite{Fort98,Leuenberger99}, the role of the nuclear
spins is less important. Although the increased mass of the isotopes
changes the spin--phonon coupling, this effect seems to be small.

We can also exclude that the change of mass for the three isotopic samples
has induced a significant change in the magnetic anisotropy of the
clusters. In fact the measurements below $T <$ 0.35~K,
where spin--phonon coupling is negligible, have shown that
(i) relative positions of the resonances as a function of the
longitudinal field $H_z$ are unchanged~\cite{remark2}, and (ii) all three
samples have the same period of oscillation of $\Delta$ as a function of
the transverse field $H_x$~\cite{Science99}, a period which is
very sensitive to any change of the anisotropy constants.

In conclusion, we presented detailed measurements 
based the Landau Zener method which demonstrated again that molecular magnets 
offer an unique opportunity to explore the quantum dynamics of a large but finite spin.
We believe that a more sophisticated theory is needed which describes the
dephasing effects of the environment.

\stars

D. Rovai, and C. Sangregorio are acknowledged for help by sample preparation. 
We are indebted to J. Villain for many fruitful discussions. 
This work has been supported by DRET and Rhone-Alpe.

\newpage

\begin{thebibliography}{99}
\bibitem{remark_Loss} Leuenberger and Loss \cite{Leuenberger00} 
proposed recently that the 
incoherent Landau--Zener transition probability $P_{\rm inc}$ 
has an exponent that is 
twice as large as the one of the coherent Landau--Zener transition 
probability $P_{\rm coh}$. This is simple due to a different definition. 
Whereas $P_{\rm coh}$ is defined as an escape probability, 
$1 - P_{\rm inc}$ is defined as the difference of population of up and down spins.
In all previous publications \cite{Science99,WW_PhD}, 
we did take into account this factor two 
when converting the measured change of magnetisation into 
an escape probability of incoherent Landau--Zener transitions (Eq. \ref{eq_LZ}).
\bibitem{remark_levels} Phonon induced transitions with $|\Delta M| > 2$
are very small \cite{Fort98}. Also the Boltzmann factor $e^{-E_{10,M}/k_BT}$
is very small for $M < 8$ and $T < 1$ K.
\bibitem{remark_const} Using Eq.~\ref{eq_H_biax} we found 
$\Delta_{-9,9}/k_B = 0.5 \times 10^{-5}$ K, 
$\Delta_{-8,8}/k_B = 0.2 \times 10^{-3}$ K, 
$E_{10,9}/k_B$ = 5.3 K, and 
$E_{10,8}/k_B$ = 9.8 K for $\vec{H} = 0$.
\bibitem{remark2} We observed a small shift of the resonances of the 
order of magnitude of 1~mT, positive for $^{57}$Fe$_8$ and negative for
$^{\rm D}$Fe$_8$ ($M_{\rm init}  = -M_{\rm s}$). 
This can also be attributed to the modified 
hyperfine fields. However, a quantitative measurement is complicated 
by the fact that it is impossible to have two crystal with exactly 
the same shape, i.e. the same internal fields.
\end{thebibliography}
\end{document}